\documentclass[10pt,twocolumn,letterpaper]{article}

\usepackage{cvpr}              % To produce the CAMERA-READY version
\usepackage{graphicx}
\usepackage{amsmath}
\usepackage{amssymb}
\usepackage{booktabs}

\usepackage{epsfig}
\usepackage{amsmath}
\usepackage{amssymb}
\usepackage{multirow}
\usepackage{makecell}
\usepackage{algorithmic}
\usepackage{array}

\usepackage[pagebackref,breaklinks,colorlinks,citecolor=cyan]{hyperref}

\usepackage[capitalize]{cleveref}
\crefname{section}{Sec.}{Secs.}
\Crefname{section}{Section}{Sections}
\Crefname{table}{Table}{Tables}
\crefname{table}{Tab.}{Tabs.}

\def\cvprPaperID{*****} % *** Enter the CVPR Paper ID here
\def\confName{CVPR}
\def\confYear{2022}

\begin{document}

\title{MT-TransUNet: Mediating Multi-Task Tokens in Transformers \\ for Skin Lesion Segmentation and Classification}

\author{Jingye Chen$^{1}$, Jieneng Chen$^{2}$, Zongwei Zhou$^{2}$, Bin Li$^{1}$\protect\thanks{Corresponding author}, Alan Yuille$^{2}$, Yongyi Lu$^{2 *}$\\
$^{1}$School of Computer Science, Fudan University \quad $^{2}$Johns Hopkins University \\ 
{\tt\small \{jingyechen19, libin\}@fudan.edu.cn, yylu1989@gmail.com}}


\maketitle

%%%%%%%%% ABSTRACT
\begin{abstract}
   Recent advances in automated skin cancer diagnosis have yielded performance on par with board-certified dermatologists.
However, these approaches formulated skin cancer diagnosis as a simple classification task, dismissing the potential benefit from lesion segmentation.
We argue that an accurate lesion segmentation can supplement the classification task with additive lesion information, such as asymmetry, border, intensity, and physical size; in turn, a faithful lesion classification can support the segmentation task with discriminant lesion features.
To this end, this paper proposes a new multi-task framework, named MT-TransUNet, which is capable of segmenting and classifying skin lesions collaboratively by mediating multi-task tokens in Transformers.
Furthermore, we have introduced dual-task and attended region consistency losses to take advantage of those images without pixel-level annotation, ensuring the model's robustness when it encounters the same image with an account of augmentation.
Our MT-TransUNet exceeds the previous state of the art for lesion segmentation and classification tasks in ISIC-2017 and PH2; more importantly, it preserves compelling computational efficiency regarding model parameters (48M~vs.~130M) and inference speed (0.17s~vs.~2.02s per image). Code will be available at \href{https://github.com/JingyeChen/MT-TransUNet}{https://github.com/JingyeChen/MT-TransUNet}.
\end{abstract}

\section{Introduction}
Over the past decades, skin cancer has emerged as a pressing challenge in public health, accountable for 5.4 million new cases in the United States~\cite{rogers2015incidence,sarker2018slsdeep}. 
Melanoma, the most serious type of skin cancer, holds a mortality rate of 75\%~\cite{li2018dense}. 
Due to the dauntingly high incidence and mortality rates, early detection and prevention of skin cancer are critical.
In response, recent studies have developed automated skin cancer diagnosis approaches, which achieved performance on par with board-certified dermatologists~\cite{esteva2017dermatologist,haenssle2018man}.
However, these approaches formulated skin cancer diagnosis as a simple classification task, dismissing the potential benefit from lesion segmentation.
In essence, the category of a skin lesion is determined by its asymmetry, border, intensity, and physical size~\cite{sharmeela2017classification}, which rely heavily on accurate lesion segmentation results.
A faithful classification, on the other hand, can also serve as crucial guidance of lesion segmentation by extracting discriminant lesion features from the dermoscopic image.
Naturally, in this paper, we seek to answer this critical question: 
\emph{How to beneficially integrate the task of segmentation with classification for skin cancer diagnosis?}

To address this question, we propose a single generic model that jointly learns classification and segmentation tasks in skin images. Our framework, called \textbf{M}ulti-\textbf{T}ask \textbf{TransUNet} (\textbf{MT-TransUNet)}, inherits the merits of both Convolutional Neural Networks and Transformers to capture both local details (\eg skin lesion color, texture) and long-range context (\eg skin lesion shape, physical size) in a multi-task learning pipeline. What makes our framework most distinct from the latest medical transformers (\eg \cite{chen2021transunet}) is that instead of tailoring for different downstream tasks separately, we jointly learn the two complementary tasks with the new design of classification and segmentation tokens. To mediate those two types of tokens, consistencies are further enforced in different levels, \ie the intermediate activated attention map as well as the final segmentation outputs, which we find is crucial for successful multi-task learning in skin images.

In contrast to the existing approaches, MT-TransUNet offers the following \textbf{three unique advantages}. 
(1) Aggregating long-range dependencies in the image. Unlike conventional CNN architectures, the self-attention layers in the Transformer architecture are capable of capturing long-range dependencies~\cite{dosovitskiy2020image,chen2021transunet}, which shed new light on identifying larger skin lesions from dermoscopic images.
(2) Exploiting both pixel- and image-level annotation. Although Y-Net~\cite{mehta2018net} attempted to combine segmentation and classification back in 2018, there was little performance gain owing to the asymmetrical learning objectives of the two tasks. We have overcome this barrier by regulating the internal consistency between the class attention map and lesion segmentation map.
(3) Demonstrating superior computational efficiency. The previous state of the art, MB-DCNN~\cite{xie2020mutual}, was extremely time-consuming and resource-intensive for training and testing, as the three modules in their architecture must be trained individually. Besides outperforming MB-DCNN by a large margin, our MT-TransUNet also presents a remarkable improvement in efficiency thanks to the design of a shared encoder.

We validate the effectiveness of MT-TransUNet on three public datasets, \ie ISIC-2017, ISIC Additional, and PH2 datasets.
Our experiments show that: 
(1) Combining segmentation and classification is capable of boosting the performance for each task.
(2) Regulating the internal consistency between the class attention map and lesion segmentation map can mediate the learning objectives across classification and segmentation.
(3) MT-TransUNet is more data-efficient and model-efficient in training and testing than prior arts. 
These results are attributable to the following key observation: 
\textit{Skin lesion classification and segmentation share a similar goal---recognizing lesions from the image and distinguishing lesion categories---thus, it is advantageous to train them jointly with a shared encoder network.}

To summarize, our contribution is three-fold:

\begin{enumerate}
\item We propose a single generic multi-task framework, named MT-TransUNet, that segment and classify skin lesions simultaneously by exploiting the potential of dermoscopy images with either pixel-level or image-level annotation.

\item We introduce dual-task and attended region consistency losses for mediating the classification and segmentation heads without pixel-level annotation, enhancing the robustness of the model when it encounters the same image undergoing various data augmentation.

\item Our MT-TransUNet exceeds the previous state of the art~\cite{xie2020mutual} for both segmentation and classification tasks, and more importantly, preserving compelling computational efficiency regarding model parameters (48M~vs.~130M) and inference speed (0.17s~vs.~2.02s per image).
\end{enumerate}

\section{Related Works}

\subsection{Skin Image Analysis}
\paragraph{Skin Lesion Segmentation:} Before the deep learning era, most methods tend to leverage traditional strategies, including thresholding, active contour models, and clustering \cite{hemalatha2018active,ravichandran2009color,yogarajah2010dynamic}. 
In the period of the deep learning era, deep neural networks gradually take the place of traditional feature extraction strategies, embracing an end-to-end manner to tackle skin image segmentation. \cite{nasr2017dense} puts forward a 19-layer neural network to segment skin images in the absence of prior knowledge of given datasets. 
\cite{mirikharaji2018star} attempts to incorporate the prior knowledge about the structure of target objects. The authors encode the star shape prior with a novel loss function, which penalizes the non-star shape predictions. \cite{sarker2018slsdeep} proposes an end point error to address the confusion of segmentation predictions around boundaries. 
\cite{liu2021skin} simultaneously predicts the segmentation mask and edge map to boost the performance. \cite{dong2021fac} leverage the feedback fusion block to concatenate the latter features with former ones to enable multiplexing of parameters.

\begin{figure*}[t]
    \centering
    \includegraphics[width=0.98\textwidth]{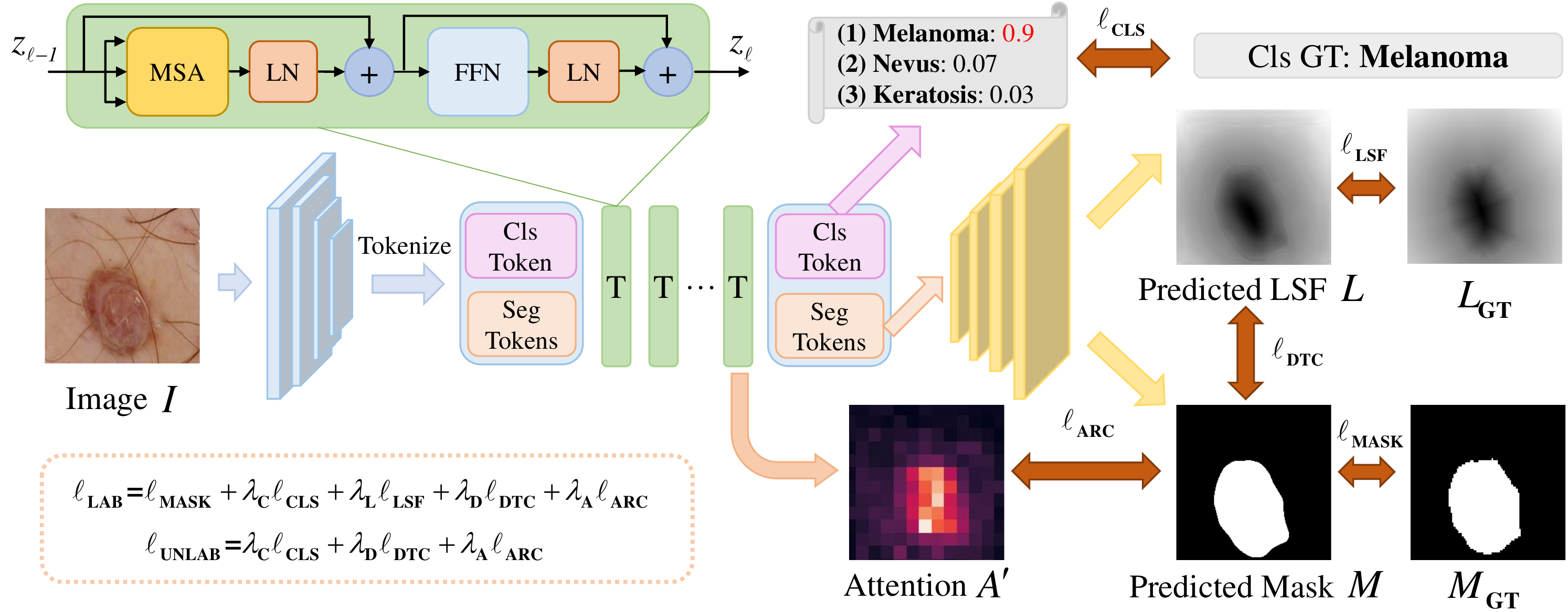}
    \caption{The overall architecture of MT-TransUNet. It mainly comprises ``cls'' and ``seg'' tokens to tackle the tasks of classification and segmentation. Dual-task consistency is introduced to exploit the images without segmentation ground truth. We also put forward an attended region consistency between segmentation and classification heads. ``T'', ``LN'', ``MSA'', and ``FFN'' denote Transformer layer, Layer Normalization, Multi-head Self-Attention, Feed-Foward Network, respectively. $\ell_{\text{LAB}}$ and $\ell_{\text{UNLAB}}$ denote the supervision for dataset with or without segmentation masks.}
    \label{fig:architecture}
\end{figure*}

\paragraph{Skin Image Classification:} Traditionally, people usually rely on handcrafted features to train a classifier, including support vector machine~\cite{alquran2017melanoma}, K-nearest-neighbor~\cite{narayanan2017automatic}, etc.
However, the design of handcrafted features is time-consuming and such low-capacity features hamper the improvement of classification performance. 
With the flourishing development of deep learning, recently many works are inclined to leverage deep neural networks for designing classifiers. For example,  \cite{zhang2018skin} proposes attention residual learning to improve the ability of discriminative representation. \cite{hagerty2019deep} takes advantage of both hand-crafted and deep features extracted by ResNet to construct a melanoma recognizer. 

\paragraph{Multi-task Learning:} Compared with those works that treat segmentation and classification tasks separately, there are few works dealing with both tasks in one model. \cite{yu2016automated} uses a two-stage framework that first segments the skin regions, which are sequentially cropped for recognition. \cite{gonzalez2018dermaknet} leverage the predicted segmentation masks to train a better classify.  Actually, both of them ignore the benefits that classifications bring to segmentation. MB-DCNN~\cite{xie2020mutual} uses a mutual bootstrapping manner to train segmentation and classification networks. Nevertheless, features between two tasks can not be shared because the two tasks are not trained in an end-to-end manner.

\subsection{Vision Transformer}

Transformer is firstly proposed to address problems concerning natural language processing. It mainly consists of a self-attention module to capture long-range dependencies and feed-forward module to project features to new latent space. Recently, ViT~\cite{dosovitskiy2020image} is proposed to tackle nature image recognition tasks based on Transformer. It first splits the images into several non-overlapping patches, then utilizes Transformer to calculate global information between each token. An additional token is appended for recognition tasks. Inspired by this, many researchers try to leverage this fashion for many purposes, such as TransUNet~\cite{chen2021transunet}, TransReID~\cite{he2021transreid}, etc. 
% Benefiting from long-range awareness brought by self-attention modules, these works perform better than previous ones. 
Furthermore, there are some variations of Transformer, like Swin Transformer~\cite{liu2021swin} that employs shifted windows to calculate local self-attention, and PVT~\cite{wang2021pyramid} that combines the fashion of pyramid network with Transformer to capture features from multiple stages.    

\subsection{Consistency Regularization}
The consistency regularization is widely used in semi-supervised learning. For example, \cite{tarvainen2017mean} designs a teacher-student consistency to take advantage of the dataset without segmentation labels. In detail, the student model is optimized by the consistency loss with regard to the targets predicted by the teacher model. \cite{luo2020semi} takes advantage of the consistency loss to boost the detection performance. \cite{zamir2020robust} put forward cross-task consistency based on inference-path invariance. \cite{luo2020semi} introduces the prediction of level set function as additional tasks and construct dual-task consistency.

\section{Method}
    In this section, we first introduce the basic design of MT-TransUNet that incorporates segmentation and classification tasks in one model, followed by the dual-task consistency (DTC) which is designed for leveraging unlabeled datasets. Then we further introduce the attended region consistency (ARC) between segmentation and classification heads. Finally, the overall loss function is presented.

\subsection{Multi-Task TransUNet}
The single-task TransUNet~\cite{chen2021transunet} follows the basic design of UNet~\cite{ronneberger2015u} that utilizes skip connections between corresponding layers to enhance local details in feature maps. Due to the intrinsic locality in convolution, UNet is weak to model long-range dependencies, thus subpar to tackle situations such as segmenting large regions in skin images. Under this circumstance, TransUNet introduces a strong encoder by incorporating several Transformer layers during feature extraction (refer to green rectangles in Figure \ref{fig:architecture}). Specifically, given a skin image $I \in \mathbb{R}^{H \times W \times C}$, the encoder first downsamples $I$ for four times using ResNet50~\cite{he2016deep} and generate a feature map $F \in \mathbb{R}^{H^{\prime} \times W^{\prime} \times C^{\prime}}$. The generated feature map is further split into $N$ non-overlapping patches  $\{\mathbf{x}^{i}_{s} \in \mathbb{R}^{P^{2} \cdot C^{\prime}} | i=1,2,...,N\}$ as segmentation tokens, each of which is of size $P \times P$ and $N = \frac{H^{\prime}W^{\prime}}{P^{2}}$. Here we append an additional zero-initialized classification token $\mathbf{x}_{c}$ to build up a multi-task framework. Each of the tokens is mapped to a latent $D$-Dimensional embedding space through a learnable linear projection $\mathbf{E}$. Different from the original implementation of TransUNet, we discard the positional embedding for the convenience of multi-scale inputs during testing. We highlight that the deep backbone can obtain the positional cues through padding~\cite{islam2020much}. Hence, the list $\mathbf{z}_{0}$ of embedded tokens are as follows:

\begin{equation}
\label{equa:linear projection}
\mathbf{z}_{0} = [\mathbf{x}^{1}_{s}\mathbf{E}; \mathbf{x}^{2}_{s}\mathbf{E}; \cdots; \mathbf{x}^{N}_{s}\mathbf{E}; \mathbf{x}_{c}\mathbf{E}]
\end{equation}

The Transformer layer mainly consists of Multihead Self-Attention (MSA) to correlate each token through long-range dependencies, as well as Feed-Forward Network (FFN) to project features to new latent space. The outputs of the $l$-layer are shown as folows:

\begin{equation}
\label{equa:MSA}
\mathbf{z}^{\prime}_{l} = \text{MSA}(\text{LN}(\mathbf{z}_{l-1})) + \mathbf{z}_{l-1}
\end{equation}

\begin{equation}
\label{equa:FFN}
\mathbf{z}_{l} = \text{FFN}(\text{LN}(\mathbf{z}^{\prime}_{l})) + \mathbf{z}^{\prime}_{l}
\end{equation}

After being enhanced by $n$ successive Transformer layers, the generated tokens are separated into two parts. For \textbf{classification} tokens, we simply use a fully connected layer to reduce the dimension of features (equal to the number of categories), then employ a cross-entropy loss $\ell_{\text{CLS}}$ for supervision. For \textbf{segmentation} tokens, they are upsampled four times in a cascade manner with bilinear interpolation and two convolutional neural networks. Specifically, in the first three steps, the feature maps of the same size in the ResNet50 encoder are concatenated in feature channels using skip connection to enhance the representation. Finally, we upsample the last feature map, resulting in two predictions, including a segmentation mask $M$ and a level set function $L$, both of which are of spatial size $H \times W$. Please note that the level set function is introduced for conducting dual-task consistency (describe in the next section in detail). In detail, the segmentation masks distinguish foregrounds and backgrounds, while the level set functions capture geometric active contour and distance information. The level set function is calculated as follows:

\begin{equation}
\label{equa:level set function}
\mathcal{T}(x)=\left\{\begin{array}{lc}-\inf _{y \in \partial S}\|x-y\|_{2}, & x \in \mathcal{S}_{\text {in }} \\ 0, & x \in \partial \mathcal{S} \\ +\inf _{y \in \partial S}\|x-y\|_{2}, & x \in \mathcal{S}_{\text {out }}\end{array}\right.
\end{equation}

Specifically, the generated segmentation mask is supervised by a cross-entropy loss, and the predicted level set function is supervised by an L2 loss.
\begin{equation}
\label{equa:mask loss}
\ell_{\text{MASK}} = \text{CrossEntropyLoss}(M, M_{\text{GT}})
\end{equation}
\begin{equation}
\label{equa:lsf loss}
\ell_{\text{LSF}} = \text{L2}(L, L_{\text{GT}})
\end{equation}
 where $M_{\text{GT}}$ and $L_{\text{GT}}$ are ground truth of segmentation mask and level set function, respectively. 

In this manner, the predictions of segmentation and classification can be generated at the same time, which distinguishes it from MB-DCNN~\cite{xie2020mutual} that employs multi-bootstrapping fashion which is time-consuming.

\begin{figure}[t]
    \centering
    \includegraphics[width=0.40\textwidth]{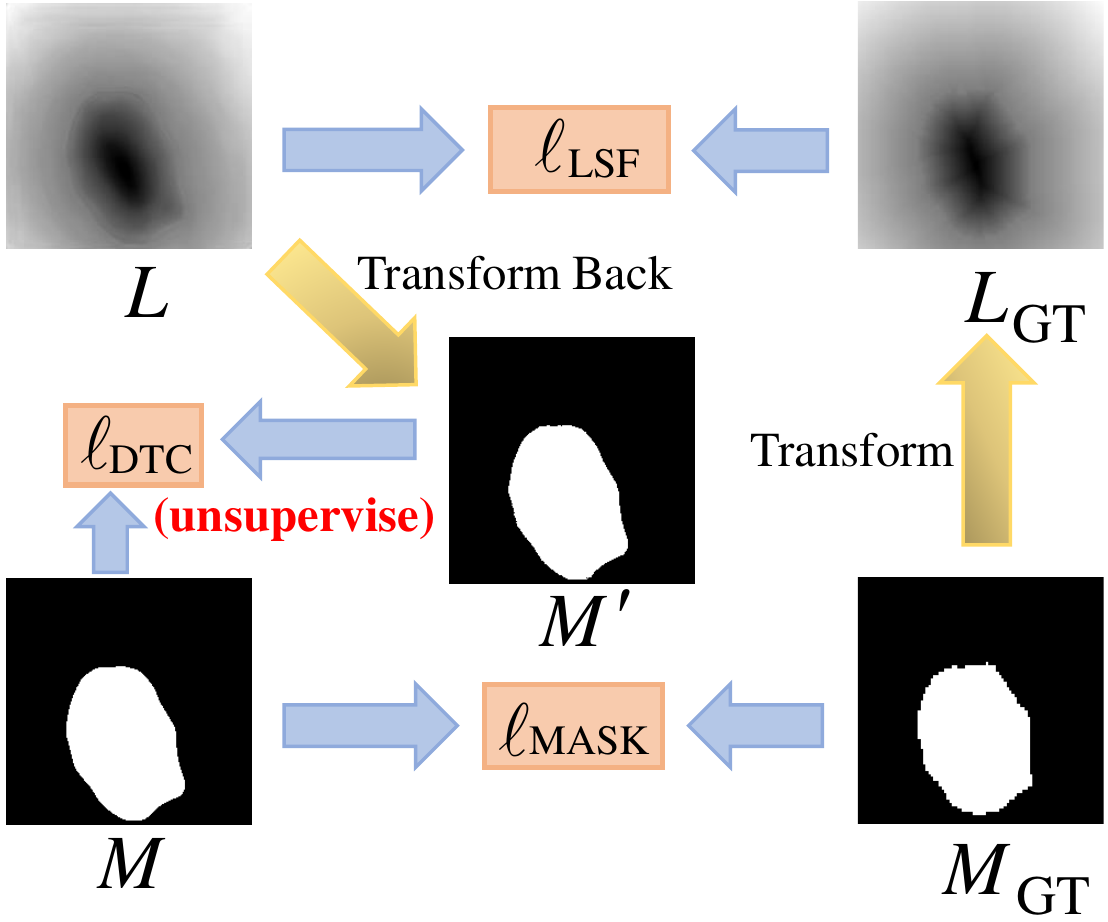}
    \caption{Dual-task consistency. For dataset without segmentation labels, the predicted level set function is first transformed back to a mask presentation, then conduct a consistency loss with the predicted segmentation mask.}
    \label{fig:dual_task_consistency}
\end{figure}

\subsection{Dual-Task Consistency}
Generally, annotating pixel-level segmentation labels is time-consuming compared with annotating classification labels. Moreover, we observe that there is an additional set only designed for classification tasks consisting of 1,320 skin images, which are widely used in the previous methods~\cite{xie2020mutual,xie2019semi} to boost the performance of classification. On the contrary, segmentation labels are not available in this part, which puts obstacles on training such a multi-task framework. One naive solution is that we only need to optimize the classification branch when training on the additional part. However, we suggest that the additional dataset can indeed provide cues for the segmentation tasks by leveraging dual-task consistency between segmentation mask $M$ and level set function $L$. Empirically, a robust model is capable of achieving high consistency between correlative predictions~\cite{zamir2020robust}.

As is demonstrated in Figure \ref{fig:dual_task_consistency}, we first multiply the generated level set function with $k$ (a large value), and employ the sigmoid function to transform it back to the mask representation. The calculation is shown as follows:

\begin{equation}
\label{equa:transform back}
M^{\prime} = \text{Sigmoid}(k \cdot L)
\end{equation}

To formulate the dual-task consistency, we utilize an L2 loss to constrain two masks.

\begin{equation}
\label{equa:two masks}
\ell_{\text{DTC}} = \text{L2}(M, M^{\prime})
\end{equation}

\begin{figure}[t]
    \centering
    \includegraphics[width=0.47\textwidth]{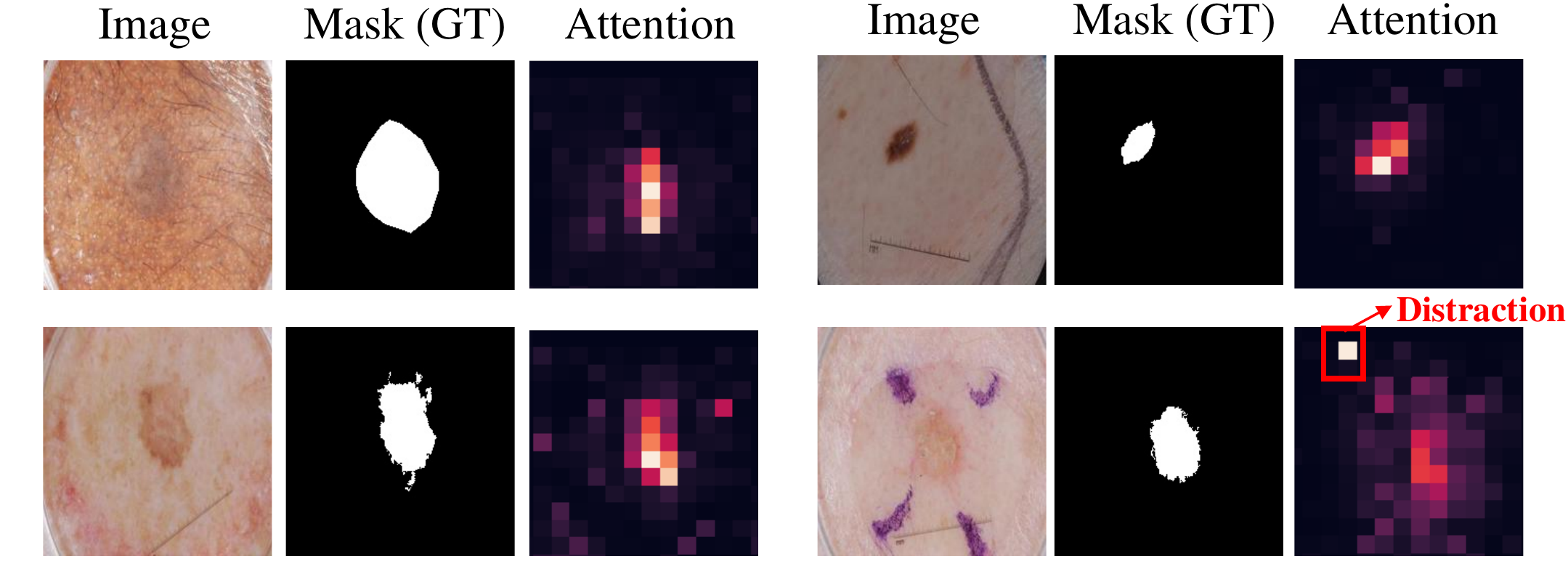}
    \caption{The attention maps of classification tokens and segmentation results are mainly focus on the foregrounds. We propose the attended region consistency loss to address the distraction phenomenon in attention maps.}
    \label{fig:attended regions}
\end{figure}

\subsection{Attended Region Consistency}
When training MT-TransUNet, we observe an interesting phenomenon: two tasks are inclined to attend to similar regions like foregrounds in skin images (see Figure \ref{fig:attended regions}). However, there are some bad cases where classification tokens attempt to attend to regions of background. Through visualization, we find that the classification branch is sensitive to hair or artificial objects (like rulers and scissors), which will cause a distraction problem. According to the ABCD rules~\cite{kasmi2016classification}, it is the \textit{Asymmetry, Border, Color, Diameter} that determines the type of skin cancer, thus the distraction problem should be penalized.

Here we introduce an attended region consistency loss between predicted segmentation masks and the generated attention maps of classification tokens. We first average the attention maps in the head channel, then we denote the attention maps as $A = (a_{s}^{1}, a_{s}^{2}, ..., a_{s}^{P^{2}}, a_{c})$, where $\sum_{i=0}^{P^{2}} a_{s}^{i} + a_{c} = 1$. We discard $a^{c}$ and normalize the attention values of segmentation tokens as follows:

\begin{equation}
\label{equa:normalize}
A^{\prime} = (a_{s}^{1}, a_{s}^{2}, ..., a_{s}^{P^{2}})\quad, \quad a^{i}_{s} \gets a^{i}_{s} \bigg/ \sum_{i=0}^{P^{2}} a_{i}^{s}
\end{equation}

We also downsample the predicted mask $M$ to $M^{\prime} \in \mathbb{R}^{ P \times P }$ with bilinear interpolation. Furthermore, we design the attended region consistency loss as follows:

\begin{equation}
\label{equa:attended region consistency}
\ell_{\text{ARC}} = (1 - M^{\prime}) \cdot A^{\prime}
\end{equation}

Specifically, lower $\ell_{\text{ARC}}$ indicates that the attended regions are consistent with predicted foregrounds in the segmentation masks.   

\subsection{Overall Loss Function}
Our MT-TransUNet is capable of simultaneously tackling datasets with and without segmentation masks. In detail, the datasets with segmentation labels are supervised as:

\begin{equation}
\label{equa:label data supervised}
\ell_{\text{LAB}} =  \ell_{\text{MASK}} + \lambda_{\text{C}}\ell_{\text{CLS}} + \lambda_{\text{L}}\ell_{\text{LSF}} + \lambda_{\text{D}}\ell_{\text{DTC}}  + \lambda_{\text{A}}\ell_{\text{ARC}}
\end{equation}
and those without segmentation labels are supervised as:
\begin{equation}
\label{equa:unlabel data supervised}
\ell_{\text{UNLAB}} = \lambda_{\text{C}}\ell_{\text{CLS}} + \lambda_{\text{D}}\ell_{\text{DTC}} + \lambda_{\text{A}}\ell_{\text{ARC}}
\end{equation}

\section{Experiments}
In this section, we firstly describe the datasets, evaluation metrics, as well as the implementation details. Then we conduct some ablation studies to verify the effectiveness of each component in the proposed architecture. 

\subsection{Dataset}

\noindent\textbf{ISIC-2017 Dataset} contains 2000 dermoscopic images for training, 150 for validation, and 600 for testing \footnote{https:\/\/challenge.isic-archive.com\/landing\/2017}. Each image is paired with a segmentation ground truth (foreground and background) and a classification ground truth (melanoma, nevus, and seborrheic keratosis).

\medskip\noindent\textbf{ISIC Additional Dataset} involves 1320 dermoscopic images only paired with classification labels. It derives from ISIC archive\footnote{https:\/\/www.isic-archive.com\/}, which is the largest publicly available collection of skin lesions.

\medskip\noindent\textbf{PH2 Dataset}~\cite{mendoncya2013dermoscopic} comprises 200 dermoscopic images, including 160 nevus as well as 40 melanomas. Both segmentation and classification labels are available. It only contains two categories, including melanoma and nevus.

\subsection{Evaluation Metrics}
For the \textbf{segmentation} results, we employ five common evaluation metrics for validation, including Jacarrd score (JA), Dice score (DI), pixel accuracy (pixel-AC), pixel-sensitivity (pixel-SE), and pixel-specitivity (pixel-SP). For the \textbf{classification} results, we use four common evaluation metrics to verify the performance of the proposed method, including accuracy (AC), sensitivity (SE), specitivity (SP), and Area Under Curve (AUC). The detailed criteria are defined following~\cite{xie2020mutual}.

\begin{table}[t]
\centering
\centering
\scalebox{0.75}{
\begin{tabular}{l||c|c|c|c}
\hline
\centering Setting  &  JA  & DI & M\_ACC & K\_ACC   \\ \hline
(1) $\ell_{\text{MASK}}$ &  79.1 & 87.0 & - & -   \\ \hline
(2) $\ell_{\text{CLS}}$ &   - & - & 88.2 & 92.5  \\ \hline
(3) $\ell_{\text{MASK}}$ + $\ell_{\text{CLS}}$ &  78.1 & 86.5 & 88.0 & 92.2  \\ \hline 
% (4) & \checkmark && \checkmark &&& 79.2 & \textbf{87.3} & - & - & - & -  \\ \hline
(4) $\ell_{\text{MASK}}$ + $\ell_{\text{CLS}}$ + $\ell_{\text{LSF}}$ &  79.4 & 87.2 & 88.7 & 92.5 \\ \hline
(5) $\ell_{\text{MASK}}$ + $\ell_{\text{CLS}}$ + $\ell_{\text{LSF}}$ + $\ell_{\text{DTC}}$ &   79.3 & \textbf{87.3} & 88.7 & \textbf{93.0}  \\ \hline
(6) $\ell_{\text{MASK}}$ + $\ell_{\text{CLS}}$ + $\ell_{\text{LSF}}$ + $\ell_{\text{DTC}}$ + $\ell_{\text{ARC}}$ &   \textbf{79.5} & \textbf{87.3} & \textbf{89.0} & \textbf{93.0} \\ \hline
\end{tabular}}
\caption{Ablation studies on multi-task framework,  dual-task consistency, and attended region consistency.}
\label{tab:ablation studies}
\end{table}

\begin{table}[t]
\centering
\centering
\scalebox{0.83}{
\begin{tabular}{p{1.5cm}<{\centering}|| p{1.5cm}<{\centering} | p{1.5cm}<{\centering} | p{1.5cm}<{\centering} | p{1.5cm}<{\centering}}
\hline
$n$ Layer & JA &  DI & M\_ACC & K\_ACC \\ \hline
0 & 79.4 & 87.1 & - & - \\ \hline
2 & 79.2 & 86.9 & 87.6 & 90.3 \\ \hline
4 & \textbf{79.5} & \textbf{87.3} & \textbf{89.0} & \textbf{93.0} \\ \hline
6 & 79.4 & \textbf{87.3} & 88.6 & 91.5 \\ \hline
8 & 79.2 & 87.1 & 88.3 & 89.7 \\ \hline
\end{tabular}}
\caption{Ablation studies on the number of Transformer layers. The performance reaches the best when $n=4$.}
\label{tab:transformer layer}
\end{table}

Please note that we follow the ISIC-2017 competition to divide the classification task into two subtasks, including melanoma classification and seborrheic keratosis classification. We utilize the metrics with different prefixes for these subtasks, \eg M\_ACC for melanoma classification and K\_ACC for keratosis classification.

\subsection{Implementation Details}
The proposed method is implemented with PyTorch. All experiments are carried out on one GTX1080Ti GPU with 11GB memory. The model is trained using Adam~\cite{kingma2014adam} optimizer. We set the batch to 8. Specifically, we construct a two-stream sampler to simultaneously optimize samples with and without segmentation labels, and the batch size of each part is equally set to 4. The learning rate is set to $10^{-5}$ and it linearly decreases to 0 during 40000 iterations. The ResNet50 and Transformer layers are pre-trained with ImageNet~\cite{deng2009imagenet}. Empirically, we set $\lambda_{\text{C}}$, $\lambda_{\text{L}}$, and $\lambda_{\text{A}}$ to 0.25, 5.0, and 1.0. For $\lambda_{\text{D}}$, we utilize an exponential ramp-up with length 40 for better convergence. We set $k$ to 1500. We employ 4 Transformer layers are used to enhance features. The patch size is set to 16 $\times$ 16.

During training, we randomly flip skin images horizontally or vertically, then crop the images from the central at multiple scales for data augmentation. During testing, we leverage three different sizes of input, including $224 \times 224$, $256 \times 256$, and $288 \times 288$. For each input size, we employ horizontal and vertical flipping, as well as multi-angle rotation ($90^{\circ}, 180^{\circ}, 270^{\circ}$) to augment the original images, and finally ensemble these results as final predictions.

\begin{table*}[t]
\begin{center}
\scalebox{0.83}{
 \begin{tabular}{l|c|c|c|c|c|c|c|c|c|c}
\hline Methods & \makecell[c]{ CDNN \\ \cite{yuan2017improving} } & \makecell[c]{DDN \\ \cite{li2018dense} }  & \makecell[c]{FCN+SSP\\\cite{mirikharaji2018star}} & \makecell[c]{SLSDeep\\\cite{sarker2018slsdeep}}  & \makecell[c]{Swin-Tiny\\\cite{liu2021swin}}  & \makecell[c]{Swin-Base\\\cite{liu2021swin} } & \makecell[c]{Segmenter\\\cite{strudel2021segmenter} } & \makecell[c]{CCL+MSFA\\\cite{liu2021skin}} & \makecell[c]{MB-DCNN\\\cite{xie2020mutual}}  & MT-TransUNet   \\
\hline JA & $76.5$ & $76.5$ & $77.3$ & $78.2$ & 75.7  & 74.8  & 75.6  & 79.5 & 79.4 / 80.4  & 79.5 / \textbf{80.7} \\
\hline DI & $84.9$ & $86.6$ & $85.7$ & 87.8  & 86.2 & 85.6  & 86.1 & 87.1 & 87.0 / 87.8 &  87.3 / \textbf{88.0}  \\
\hline pixel-AC & $93.4$ & $93.9$ & $93.8$ & $93.6$  & 92.4  & 92.8 &  93.8 & 94.3 & 94.3 / 94.7 &  94.6 / \textbf{94.9}  \\
\hline pixel-SE & $82.5$ & $82.5$ & $85.5$ & $81.6$  & 86.6 & 87.1 &  83.3  & \textbf{88.8} & 87.3 / 87.4 & 88.0 / 88.2 \\
\hline pixel-SP & $97.5$ & \textbf{98.4} & $97.3$ & 98.3  & 96.9 &  96.5 & 97.2 & 96.5 & 96.4 / 96.8 & 96.5 / 96.4  \\
\hline
\end{tabular}}
\end{center}
\caption{Experimental results on the segmentation task on ISIC-2017. Since MB-DCNN~\cite{xie2020mutual} employs five-model ensembling when testing, we also follow this setting for a fair comparison (the ensembling results are shown after the slash). Our MT-TransUNet achieves the state-of-the-art performance no matter the model ensembling strategy is used or not.}
\label{tab:segmentation}
\end{table*}

%%%%%%% 分类

\begin{table*}[t]
\begin{center}
\scalebox{0.99}{
\begin{tabular}{c|c|c|c|c|c|c|c|c|c}
\hline
\multirow{2}{*} { Methods } & \multicolumn{4}{c|} { Melanoma Classification } & \multicolumn{4}{c|} { Keratosis Classification } & \multirow{2}{*} {Average ACC } \\
\cline { 2 - 9 } & M\_ACC & M\_SE & M\_SP & M\_AUC & K\_ACC & K\_SE & K\_SP & K\_AUC & ~ \\
\hline ARL-CNN~\cite{zhang2019attention} & $85.0$ & $65.8$ & $89.6$ & $87.5$ & $86.8$ & $87.8$ & $86.7$ & $95.8$ & $85.9$  \\
\hline SSAC~\cite{xie2019semi} & $83.5$ & $55.6$ & $90.3$ & $87.3$ & $91.2$ & $88.9$ & $91.6$ & $95.9$ & $87.4$  \\
\hline SDL~\cite{zhang2019medical} & $88.8$ & $-$ & $-$ & $86.8$ & $92.5$ & $-$ & $-$ & $95.8$  & $90.7$  \\
\hline~\cite{matsunaga2017image} & $82.8$ & $\mathbf{73.5}$ & $85.1$ & $86.8$ & $80.3$ & $\mathbf{97.8}$ & $77.3$ & $95.3$ & $81.6$  \\
\hline~\cite{diaz2017incorporating} & $82.3$ & $10.3$ & $\mathbf{99.8}$ & $85.6$ & $87.5$ & $17.8$ & $\mathbf{99.8}$ & $96.5$ & 84.9 \\
\hline~\cite{menegola2017recod} & $87.2$ & $54.7$ & $95.0$ & $87.4$ & $89.5$ & $35.6$ & $99.0$ & $94.3$ & 88.4  \\
\hline~\cite{bi2017automatic} & $85.8$ & $42.7$ & $96.3$ & $87.0$ & $91.8$ & $58.9$ & $97.6$ & $92.1$  & 88.9  \\
\hline~\cite{yang2017novel} & $83.0$ & $43.6$ & $92.5$ & $83.0$ & $91.7$ & $70.0$ & $99.5$ & $94.2$ & 87.4  \\
\hline
\hline MB-DCNN~\cite{xie2020mutual} & $86.7$ & $70.1$ & $90.7$ & $89.6$ & $92.3$ & $83.3$ & $93.9$ & $96.7$ &89.5 \\
% \hline \text{MB-DCNN-en} & $87.8$ & $72.7$ & $91.5$ & $90.3$ & $93.0$ & $84.4$ & $94.5$ & $\mathbf{97.3}$ & 90.4 \\
\hline MT-TransUNet & 89.0 & $69.3$ & $91.2$ & 89.4 & 93.0 & $92.8$ & $96.3$ & $95.1$ & 91.0  \\
\hline
\hline $\text{MB-DCNN}^{*}$~\cite{xie2020mutual} & $87.8$ & $72.7$ & $91.5$ & $90.3$ & $93.0$ & $84.4$ & $94.5$ & $\mathbf{97.3}$ & 90.4 \\
\hline $\text{MT-TransUNet}^{*}$ & \textbf{89.2} & 68.0 & 92.3 & \textbf{90.6} & \textbf{93.2} & 77.6 & 97.6 & 95.7 & \textbf{91.2}  \\
\hline
\end{tabular}}
\end{center}
\caption{Experimental results of the classification task on ISIC-2017. $^{*}$ denotes five-model ensembling. Our method surpasses MB-DCNN with regard to recognition accuracy in two subtasks with or without the ensembling strategy.}
\label{tab:classification}
\end{table*}

\subsection{Ablation Studies}
Here we carry out several ablation studies concerning the joint training strategy, dual-task and attended region consistency, as well as Transformer layers. The experimental results are shown in Table \ref{tab:ablation studies} and Table \ref{tab:transformer layer}, respectively. All of the ablation studies are conducted on ISIC-2017.

\paragraph{Effectiveness of Joint Training:} Firstly, we conduct experiments with a single branch \ie only generate segmentation or classification predictions (Settings (1) and (2)). Then we jointly train segmentation and classification tasks (Setting (3)). From this ablation study, we observe that naively combining segmentation and classification branches in one model will decrease the performance for both tasks. For example, the Jaccard score decreases by 1\% and M\_ACC decreases by 0.2\%. We suppose that the reason is: For the classification task, the model attempts to learn the most discriminative features (usually high-level features) rather than exploit the whole information. However, for the segmentation task, the model focuses more on low-level features like edges or colors to judge whether a pixel belongs to foregrounds or backgrounds. Base on these observations, we attempt to boost the performance of this multi-task framework in two ways: 1) Utilize semi-supervised training to take advantage of data without segmentation masks. 2) Exploit the consistency between segmentation and classification heads. 

\paragraph{Effectiveness of Dual-Task Consistency:} Since annotating segmentation masks for skin images is time-consuming, there is a large number of skin images only with classification labels. Hence, we employ a semi-supervised manner to take advantage of the additional dataset through dual-task consistency (Settings (4) and (5)). When the level set function head is appended along with the segmentation mask branch, the performance of both tasks increases and is better than that of single branch settings \eg (1) and (2). Compared with the segmentation mask, the level set function cares more about the information of the edge, thus helping the segmentation tasks indirectly and achieving better performance than setting (1) in terms of Jaccard score. When the consistency loss is introduced in this architecture, the performance of both tasks is further boosted. For example, compared with (1), the Dice score can be boosted by 0.2\% in the segmentation task. While compared with (2), the M\_ACC can be boosted by 0.5\%.

\paragraph{Effectiveness of Attended Region Consistency:} The introduction of the dual-task mainly concentrates on the consistency of two low-level tasks. To further exploit the consistency between segmentation and classification heads, we put forward an attended region consistency. As is shown in setting (6), the performance reaches the best compared with all its counterparts. Intuitively, the attended region consistency can rectify the distraction phenomenon of the classification branch, thus resulting in higher classification performance. 

\paragraph{Effectiveness of Transformer Layers:} As is shown in Table \ref{tab:transformer layer} we conduct the ablation studies to verify the effectiveness of the appended Transformer layers. The performance reaches the best when four Transformer layers are used. We suppose that the Transformer layers are capable of correlating each grid in feature maps through the self-attention mechanism, thus achieving better scores compared with the model without any Transformer layers.

\begin{figure*}[t]
    \centering
    \includegraphics[width=1.02\textwidth]{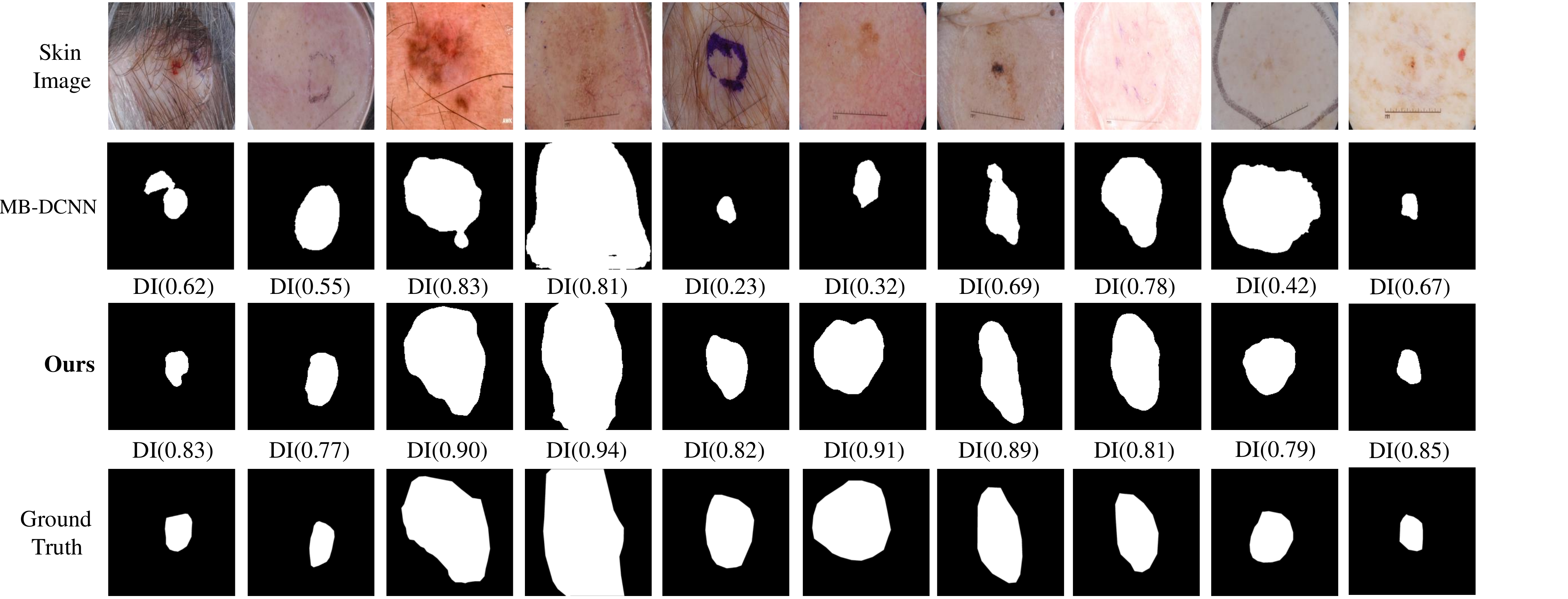}
    \caption{The segmentation predictions of our method. Compared with MB-DCNN, our method is more sensitive to the contour information, thus achieving higher performance regarding Dice score. }
    \label{fig:visualization}
\end{figure*}

\subsection{Compared with Previous Methods}
The segmentation results are shown in Table \ref{tab:segmentation}. We also conduct experiments on the existing Transformer-based architecture like Swin-Transformer~\cite{liu2021swin} and Segmenter~\cite{strudel2021segmenter}. We observe that these Transformer-based architectures are subpar on this task since they do not introduce any priors with regard to skin images to the model. Through the results, our method achieves the state-of-the-art performance in terms of Jacarrd score, pixel-AC, as well as pixel-SE, which indeed verifies the ability of the proposed method. The results of the classification task are demonstrated in Table \ref{tab:classification}. Our method achieves the best average accuracy compared with its counterparts.

\subsection{Generalization Ability}\label{section:ph2}
In this section, we follow MB-DCNN~\cite{xie2020mutual} to validate the generalization ability of the proposed method in two ways: 1) Test on the PH2 dataset using the model pre-trained on ISIC-2017 dataset and the additional dataset. 2) Use four-fold cross-validation, \textit{i.e.} regard the trained MB-DCNN as a pre-trained one, and use three folds of PH2 dataset to fine-tune the model, while testing the fine-tuned model on the other fold of the PH2 dataset. The experimental results of the segmentation and classification tasks are shown in Table \ref{tab:ph2 classification} and Table \ref{tab:segmentation}. Through the experimental results, we observe that our model is robust compared with MB-DCNN when generalized to another dataset. For example, our method is 3.1\% better in terms of recognition accuracy and 2.8\% better in terms of Jacarrd score compared with MB-DCNN~\cite{xie2020mutual} with fine-tuning. 

\begin{table}[t]
\centering
\centering
\scalebox{0.76}{
 \begin{tabular}{l|c|c|c||c|c}
\hline Datasets & \multicolumn{5}{c} { PH2 } \\
\hline Methods &  \makecell[c]{ mFCNPI\\\cite{bi2017dermoscopic}} & \makecell[c]{ RFCN\\\cite{yuan2017automatic}} & \makecell[c]{ SLIC\\\cite{patino2018automatic}}  & \makecell[c]{MB-DCNN\\\cite{xie2020mutual}}  & MT-TransUNet  \\
\hline JA & 84.0 & - & - & 86.7 / 89.4 & 88.5 / \textbf{92.2} \\
\hline DI & 90.7 & 93.8 & - & 92.6 / 94.2 & 93.6 / \textbf{95.9} \\
\hline pixel-AC & 94.2 & - & 90.4 & 95.8 / 96.4 & \textbf{96.5} / 93.5  \\
\hline pixel-SE & 94.9 & - & 91.0 & \textbf{97.9} / 96.7 & 97.2 / 96.5 \\
\hline pixel-SP & 94.0 & - & 89.7 & 95.1 /  94.6 & 95.7 / \textbf{98.0} \\
\hline
\end{tabular}}
\caption{Experimental results of the segmentation task on PH2 dataset. \textit{a/b} denotes the results of \textit{directly testing} / \textit{finetuning} settings.}
\label{tab:ph2 segmentation}
\end{table}

%%%%%%%%%%%%%%%%%%%%%%% move to supp
\begin{table}[t]
\centering
\centering
\scalebox{0.78}{
\begin{tabular}{l|c|c|c|c}
\hline Methods & M\_AC & M\_SE & M\_SP & M\_AUC \\
\hline CICS~\cite{barata2017development} & - & \textbf{100.0} & 88.2 & - \\
\hline MFLF~\cite{barata2015melanoma} & - & 98.0 & 90.0 & - \\
\hline CCS~\cite{barata2014improving} & - & 92.5 & 76.3 & 84.3 \\
\hline
\hline MB-DCNN~\cite{xie2020mutual} (test) & 88.5 & 82.5 & 90.0 & 95.7 \\
\hline Ours (test) & 95.0 & 82.5 & 98.1 & 96.0 \\
\hline
\hline MB-DCNN~\cite{xie2020mutual} (ft) & 94.0 & 95.0 & 93.8 & 97.7 \\
\hline Ours (ft) & \textbf{97.1} & 91.2 & \textbf{99.0} & \textbf{98.6} \\
\hline
\end{tabular}}
\caption{Experimental results of the classification task on PH2 dataset.}
\label{tab:ph2 classification}
\end{table}

\subsection{Model Efficiency}
In this section, we compare the efficiency of the proposed model with another multi-task framework MB-DCNN~\cite{xie2020mutual}. Specifically, MB-DCNN comprises three separate networks, including CoarseSN, MaskCN, and EnhancedSN. The total amount of parameters is 130M, which is far more than our model (48M). During training, it only takes 8 hours to train our model, which is more efficient compared with MB-DCNN (48 hours). As for the inference time, when the batch size is set to 1, our method is capable of generating both segmentation and classification results in 0.17 seconds. On the contrary, it costs MB-DCNN 2.02 seconds in total to generate both predictions. Furthermore, since the EnhancedSN relies on the intermediate results produced by MaskCN, it will cost large disk storage. Hence, our model surpasses MB-DCNN in terms of model efficiency.

\begin{figure}[t]
    \centering
    \includegraphics[width=0.45\textwidth]{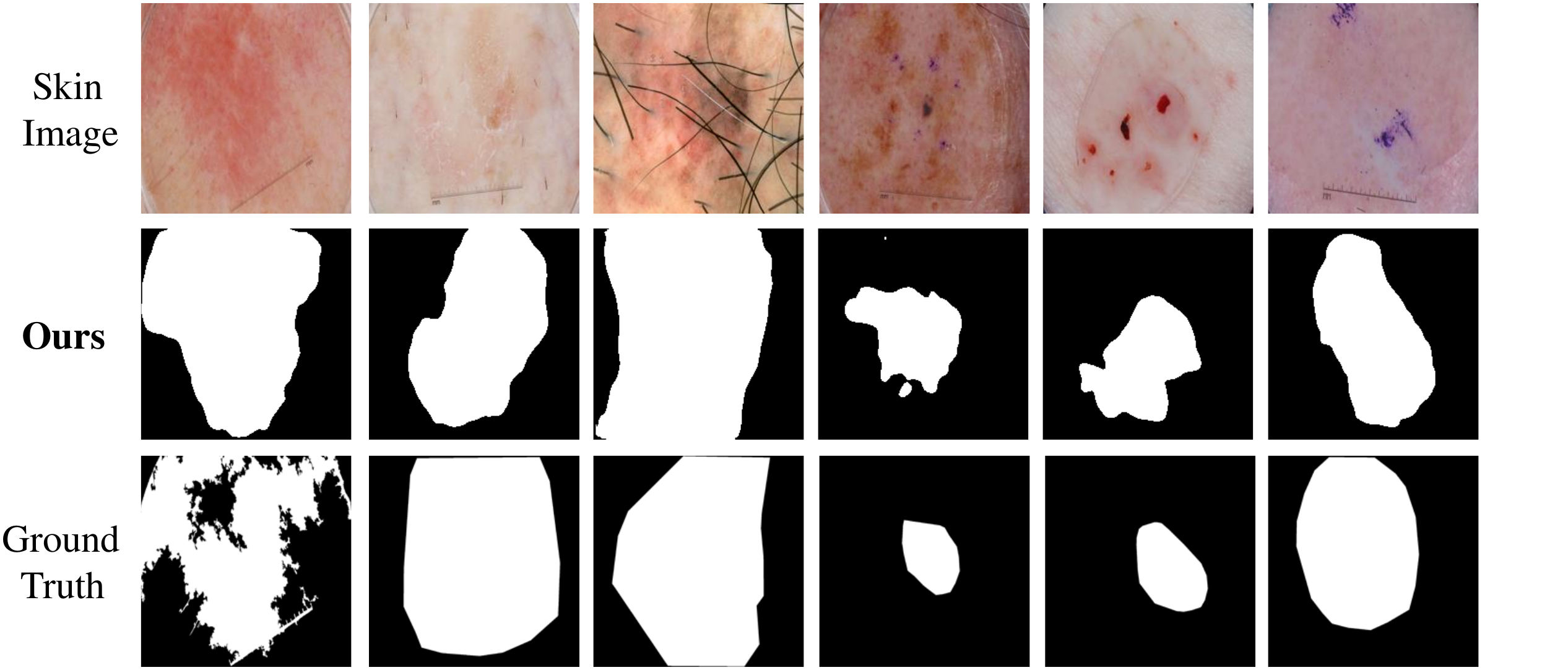}
    \caption{Some failure cases of segmentation predictions.}
    \label{fig:Failure case}
\end{figure}

\subsection{Failure Cases}
Several failure cases are shown in Figure \ref{fig:Failure case}. We observe that the low-contrast skin images still bring great difficulties for our model. Besides, when there is some occlusion in the skin image \eg hair or rulers, the performance of segmentation will also decrease. To tackle these failure cases, we suppose that more additional preprocessing strategies should be utilized to enhance the original images, such as contrast enhancement and occlusion removal.

\section{Conclusion}
In this paper, we proposed a multi-task framework called MT-TransUNet to combine segmentation and classification tasks in one model. We represent each task as corresponding tokens and leverage Transformer layers to correlate tasks in feature domains. To utilize the dataset without segmentation labels, we utilize a semi-supervised fashion by introducing the level set function as a dual-task, then constrain the consistency between the segmentation masks and the level set function. Besides, we also put forward an attended region consistency between segmentation and classification heads. The experimental results show that our method can achieve state-of-the-art performance on both tasks while preserving better efficiency compared with previous methods.

%%%%%%%%% REFERENCES
{\small
\bibliographystyle{ieee_fullname}
\bibliography{main}

\begin{thebibliography}{10}\itemsep=-1pt

\bibitem{alquran2017melanoma}
Hiam Alquran, Isam~Abu Qasmieh, Ali~Mohammad Alqudah, Sajidah Alhammouri, Esraa
  Alawneh, Ammar Abughazaleh, and Firas Hasayen.
\newblock The melanoma skin cancer detection and classification using support
  vector machine.
\newblock In {\em 2017 IEEE Jordan Conference on Applied Electrical Engineering
  and Computing Technologies (AEECT)}, pages 1--5, 2017.

\bibitem{barata2014improving}
Catarina Barata, M~Emre Celebi, and Jorge~S Marques.
\newblock Improving dermoscopy image classification using color constancy.
\newblock {\em IEEE journal of biomedical and health informatics},
  19(3):1146--1152, 2014.

\bibitem{barata2015melanoma}
Catarina Barata, M~Emre Celebi, and Jorge~S Marques.
\newblock Melanoma detection algorithm based on feature fusion.
\newblock In {\em 2015 37th Annual International Conference of the IEEE
  Engineering in Medicine and Biology Society (EMBC)}, pages 2653--2656, 2015.

\bibitem{barata2017development}
Catarina Barata, M~Emre Celebi, and Jorge~S Marques.
\newblock Development of a clinically oriented system for melanoma diagnosis.
\newblock {\em Pattern Recognition}, 69:270--285, 2017.

\bibitem{bi2017automatic}
Lei Bi, Jinman Kim, Euijoon Ahn, and Dagan Feng.
\newblock Automatic skin lesion analysis using large-scale dermoscopy images
  and deep residual networks.
\newblock {\em arXiv preprint arXiv:1703.04197}, 2017.

\bibitem{bi2017dermoscopic}
Lei Bi, Jinman Kim, Euijoon Ahn, Ashnil Kumar, Michael Fulham, and Dagan Feng.
\newblock Dermoscopic image segmentation via multistage fully convolutional
  networks.
\newblock {\em IEEE Transactions on Biomedical Engineering}, 64(9):2065--2074,
  2017.

\bibitem{chen2021transunet}
Jieneng Chen, Yongyi Lu, Qihang Yu, Xiangde Luo, Ehsan Adeli, Yan Wang, Le Lu,
  Alan~L Yuille, and Yuyin Zhou.
\newblock Transunet: Transformers make strong encoders for medical image
  segmentation.
\newblock {\em arXiv preprint arXiv:2102.04306}, 2021.

\bibitem{deng2009imagenet}
Jia Deng, Wei Dong, Richard Socher, Li-Jia Li, Kai Li, and Li Fei-Fei.
\newblock Imagenet: A large-scale hierarchical image database.
\newblock In {\em CVPR}, pages 248--255, 2009.

\bibitem{diaz2017incorporating}
Iv{\'a}n~Gonz{\'a}lez D{\'\i}az.
\newblock Incorporating the knowledge of dermatologists to convolutional neural
  networks for the diagnosis of skin lesions.
\newblock {\em arXiv preprint arXiv:1703.01976}, 2017.

\bibitem{dong2021fac}
Yuying Dong, Liejun Wang, Shuli Cheng, and Yongming Li.
\newblock Fac-net: Feedback attention network based on context encoder network
  for skin lesion segmentation.
\newblock {\em Sensors}, 21(15):5172, 2021.

\bibitem{dosovitskiy2020image}
Alexey Dosovitskiy, Lucas Beyer, Alexander Kolesnikov, Dirk Weissenborn,
  Xiaohua Zhai, Thomas Unterthiner, Mostafa Dehghani, Matthias Minderer, Georg
  Heigold, Sylvain Gelly, et~al.
\newblock An image is worth 16x16 words: Transformers for image recognition at
  scale.
\newblock {\em arXiv preprint arXiv:2010.11929}, 2020.

\bibitem{esteva2017dermatologist}
Andre Esteva, Brett Kuprel, Roberto~A Novoa, Justin Ko, Susan~M Swetter,
  Helen~M Blau, and Sebastian Thrun.
\newblock Dermatologist-level classification of skin cancer with deep neural
  networks.
\newblock {\em Nature}, 542(7639):115, 2017.

\bibitem{gonzalez2018dermaknet}
Ivan Gonzalez-Diaz.
\newblock Dermaknet: Incorporating the knowledge of dermatologists to
  convolutional neural networks for skin lesion diagnosis.
\newblock {\em IEEE journal of biomedical and health informatics},
  23(2):547--559, 2018.

\bibitem{haenssle2018man}
Holger~A Haenssle, Christine Fink, R Schneiderbauer, Ferdinand Toberer, Timo
  Buhl, A Blum, A Kalloo, A~Ben~Hadj Hassen, Luc Thomas, A Enk, et~al.
\newblock Man against machine: diagnostic performance of a deep learning
  convolutional neural network for dermoscopic melanoma recognition in
  comparison to 58 dermatologists.
\newblock {\em Annals of Oncology}, 29(8):1836--1842, 2018.

\bibitem{hagerty2019deep}
Jason~R Hagerty, R~Joe Stanley, Haidar~A Almubarak, Norsang Lama, Reda Kasmi,
  Peng Guo, Rhett~J Drugge, Harold~S Rabinovitz, Margaret Oliviero, and
  William~V Stoecker.
\newblock Deep learning and handcrafted method fusion: higher diagnostic
  accuracy for melanoma dermoscopy images.
\newblock {\em IEEE journal of biomedical and health informatics},
  23(4):1385--1391, 2019.

\bibitem{he2016deep}
Kaiming He, Xiangyu Zhang, Shaoqing Ren, and Jian Sun.
\newblock Deep residual learning for image recognition.
\newblock In {\em CVPR}, pages 770--778, 2016.

\bibitem{he2021transreid}
Shuting He, Hao Luo, Pichao Wang, Fan Wang, Hao Li, and Wei Jiang.
\newblock Transreid: Transformer-based object re-identification.
\newblock {\em arXiv preprint arXiv:2102.04378}, 2021.

\bibitem{hemalatha2018active}
RJ Hemalatha, TR Thamizhvani, A~Josephin~Arockia Dhivya, Josline~Elsa Joseph,
  Bincy Babu, and R Chandrasekaran.
\newblock Active contour based segmentation techniques for medical image
  analysis.
\newblock {\em Medical and Biological Image Analysis}, 4:17, 2018.

\bibitem{islam2020much}
Md~Amirul Islam, Sen Jia, and Neil~DB Bruce.
\newblock How much position information do convolutional neural networks
  encode?
\newblock {\em arXiv preprint arXiv:2001.08248}, 2020.

\bibitem{kasmi2016classification}
Reda Kasmi and Karim Mokrani.
\newblock Classification of malignant melanoma and benign skin lesions:
  implementation of automatic abcd rule.
\newblock {\em IET Image Processing}, 10(6):448--455, 2016.

\bibitem{kingma2014adam}
Diederik~P Kingma and Jimmy Ba.
\newblock Adam: A method for stochastic optimization.
\newblock {\em arXiv preprint arXiv:1412.6980}, 2014.

\bibitem{li2018dense}
Hang Li, Xinzi He, Feng Zhou, Zhen Yu, Dong Ni, Siping Chen, Tianfu Wang, and
  Baiying Lei.
\newblock Dense deconvolutional network for skin lesion segmentation.
\newblock {\em IEEE journal of biomedical and health informatics},
  23(2):527--537, 2018.

\bibitem{liu2021skin}
Lina Liu, Ying~Y Tsui, and Mrinal Mandal.
\newblock Skin lesion segmentation using deep learning with auxiliary task.
\newblock {\em Journal of Imaging}, 7(4):67, 2021.

\bibitem{liu2021swin}
Ze Liu, Yutong Lin, Yue Cao, Han Hu, Yixuan Wei, Zheng Zhang, Stephen Lin, and
  Baining Guo.
\newblock Swin transformer: Hierarchical vision transformer using shifted
  windows.
\newblock {\em arXiv preprint arXiv:2103.14030}, 2021.

\bibitem{luo2020semi}
Xiangde Luo, Jieneng Chen, Tao Song, and Guotai Wang.
\newblock Semi-supervised medical image segmentation through dual-task
  consistency.
\newblock In {\em AAAI}, 2021.

\bibitem{matsunaga2017image}
Kazuhisa Matsunaga, Akira Hamada, Akane Minagawa, and Hiroshi Koga.
\newblock Image classification of melanoma, nevus and seborrheic keratosis by
  deep neural network ensemble.
\newblock {\em arXiv preprint arXiv:1703.03108}, 2017.

\bibitem{mehta2018net}
Sachin Mehta, Ezgi Mercan, Jamen Bartlett, Donald Weaver, Joann~G Elmore, and
  Linda Shapiro.
\newblock Y-net: joint segmentation and classification for diagnosis of breast
  biopsy images.
\newblock In {\em MICCAI}, pages 893--901, 2018.

\bibitem{mendoncya2013dermoscopic}
T Mendonc{\"y}a, PM Ferreira, J Marques, ARS Marc{\"y}al, and J Rozeira.
\newblock A dermoscopic image database for research and benchmarking.
\newblock {\em Presentation in proceedings of PH2 IEEE EMBC}, 2013.

\bibitem{menegola2017recod}
Afonso Menegola, Julia Tavares, Michel Fornaciali, Lin~Tzy Li, Sandra Avila,
  and Eduardo Valle.
\newblock Recod titans at isic challenge 2017.
\newblock {\em arXiv preprint arXiv:1703.04819}, 2017.

\bibitem{mirikharaji2018star}
Zahra Mirikharaji and Ghassan Hamarneh.
\newblock Star shape prior in fully convolutional networks for skin lesion
  segmentation.
\newblock In {\em MICCAI}, pages 737--745, 2018.

\bibitem{narayanan2017automatic}
Hema~Rajini Narayanan.
\newblock Automatic classification of skin cancer using knn, svm and cnn.
\newblock {\em Eurasian Journal of Analytical Chemistry}, 12(1):133--138, 2017.

\bibitem{nasr2017dense}
Ebrahim Nasr-Esfahani, Shima Rafiei, Mohammad~H Jafari, Nader Karimi, James~S
  Wrobel, SM Soroushmehr, Shadrokh Samavi, and Kayvan Najarian.
\newblock Dense fully convolutional network for skin lesion segmentation.
\newblock {\em arXiv preprint arXiv:1712.10207}, 2017.

\bibitem{patino2018automatic}
Diego Pati{\~n}o, Jonathan Avenda{\~n}o, and John~W Branch.
\newblock Automatic skin lesion segmentation on dermoscopic images by the means
  of superpixel merging.
\newblock In {\em MICCAI}, pages 728--736, 2018.

\bibitem{ravichandran2009color}
KS Ravichandran and B Ananthi.
\newblock Color skin segmentation using k-means cluster.
\newblock {\em International Journal of Computational and Applied Mathematics},
  4(2):153--158, 2009.

\bibitem{rogers2015incidence}
Howard~W Rogers, Martin~A Weinstock, Steven~R Feldman, and Brett~M Coldiron.
\newblock Incidence estimate of nonmelanoma skin cancer (keratinocyte
  carcinomas) in the us population, 2012.
\newblock {\em JAMA dermatology}, 151(10):1081--1086, 2015.

\bibitem{ronneberger2015u}
Olaf Ronneberger, Philipp Fischer, and Thomas Brox.
\newblock U-net: Convolutional networks for biomedical image segmentation.
\newblock In {\em MICCAI}, pages 234--241, 2015.

\bibitem{sarker2018slsdeep}
Md~Mostafa~Kamal Sarker, Hatem~A Rashwan, Farhan Akram, Syeda~Furruka Banu,
  Adel Saleh, Vivek~Kumar Singh, Forhad~UH Chowdhury, Saddam Abdulwahab,
  Santiago Romani, Petia Radeva, et~al.
\newblock Slsdeep: Skin lesion segmentation based on dilated residual and
  pyramid pooling networks.
\newblock In {\em MICCAI}, pages 21--29, 2018.

\bibitem{sharmeela2017classification}
S Sharmeela and P Asha.
\newblock Classification of skin diseases by using back propagation neural
  network and abcd rule.
\newblock {\em Global Journal of Pure and Applied Mathematics},
  13(7):3161--3171, 2017.

\bibitem{strudel2021segmenter}
Robin Strudel, Ricardo Garcia, Ivan Laptev, and Cordelia Schmid.
\newblock Segmenter: Transformer for semantic segmentation.
\newblock {\em arXiv preprint arXiv:2105.05633}, 2021.

\bibitem{tarvainen2017mean}
Antti Tarvainen and Harri Valpola.
\newblock Mean teachers are better role models: Weight-averaged consistency
  targets improve semi-supervised deep learning results.
\newblock {\em arXiv preprint arXiv:1703.01780}, 2017.

\bibitem{wang2021pyramid}
Wenhai Wang, Enze Xie, Xiang Li, Deng-Ping Fan, Kaitao Song, Ding Liang, Tong
  Lu, Ping Luo, and Ling Shao.
\newblock Pyramid vision transformer: A versatile backbone for dense prediction
  without convolutions.
\newblock {\em arXiv preprint arXiv:2102.12122}, 2021.

\bibitem{xie2019semi}
Yutong Xie, Jianpeng Zhang, and Yong Xia.
\newblock Semi-supervised adversarial model for benign--malignant lung nodule
  classification on chest ct.
\newblock {\em Medical image analysis}, 57:237--248, 2019.

\bibitem{xie2020mutual}
Yutong Xie, Jianpeng Zhang, Yong Xia, and Chunhua Shen.
\newblock A mutual bootstrapping model for automated skin lesion segmentation
  and classification.
\newblock {\em IEEE transactions on medical imaging}, 39(7):2482--2493, 2020.

\bibitem{yang2017novel}
Xulei Yang, Zeng Zeng, Si~Yong Yeo, Colin Tan, Hong~Liang Tey, and Yi Su.
\newblock A novel multi-task deep learning model for skin lesion segmentation
  and classification.
\newblock {\em arXiv preprint arXiv:1703.01025}, 2017.

\bibitem{yogarajah2010dynamic}
Pratheepan Yogarajah, Joan Condell, Kevin Curran, Abbas Cheddad, and Paul
  McKevitt.
\newblock A dynamic threshold approach for skin segmentation in color images.
\newblock In {\em ICIP}, pages 2225--2228, 2010.

\bibitem{yu2016automated}
Lequan Yu, Hao Chen, Qi Dou, Jing Qin, and Pheng-Ann Heng.
\newblock Automated melanoma recognition in dermoscopy images via very deep
  residual networks.
\newblock {\em IEEE transactions on medical imaging}, 36(4):994--1004, 2016.

\bibitem{yuan2017automatic}
Yading Yuan, Ming Chao, and Yeh-Chi Lo.
\newblock Automatic skin lesion segmentation using deep fully convolutional
  networks with jaccard distance.
\newblock {\em IEEE transactions on medical imaging}, 36(9):1876--1886, 2017.

\bibitem{yuan2017improving}
Yading Yuan and Yeh-Chi Lo.
\newblock Improving dermoscopic image segmentation with enhanced
  convolutional-deconvolutional networks.
\newblock {\em IEEE journal of biomedical and health informatics},
  23(2):519--526, 2017.

\bibitem{zamir2020robust}
Amir~R Zamir, Alexander Sax, Nikhil Cheerla, Rohan Suri, Zhangjie Cao, Jitendra
  Malik, and Leonidas~J Guibas.
\newblock Robust learning through cross-task consistency.
\newblock In {\em CVPR}, pages 11197--11206, 2020.

\bibitem{zhang2018skin}
Jianpeng Zhang, Yutong Xie, Qi Wu, and Yong Xia.
\newblock Skin lesion classification in dermoscopy images using synergic deep
  learning.
\newblock In {\em MICCAI}, pages 12--20, 2018.

\bibitem{zhang2019medical}
Jianpeng Zhang, Yutong Xie, Qi Wu, and Yong Xia.
\newblock Medical image classification using synergic deep learning.
\newblock {\em Medical image analysis}, 54:10--19, 2019.

\bibitem{zhang2019attention}
Jianpeng Zhang, Yutong Xie, Yong Xia, and Chunhua Shen.
\newblock Attention residual learning for skin lesion classification.
\newblock {\em IEEE transactions on medical imaging}, 38(9):2092--2103, 2019.

\end{thebibliography}
}

\end{document}